\documentclass[useAMS,usenatbib]{mn2e} 
\usepackage{lscape}
\usepackage{graphicx} 
\usepackage{amssymb} 
\usepackage[usenames]{color}

\title[Fundamental parameters of 8 Am stars]{Fundamental parameters of 8 Am stars: comparing observations with theory\thanks{Based 
          on observations made with the Italian Telescopio
          Nazionale Galileo (TNG) operated on the island of La Palma by the Fundaci\'on Galileo Galilei of the 
          INAF (Istituto Nazionale di Astrofisica) at the Spanish Observatorio del Roque de los Muchachos of
          the Instituto de Astrofisica de Canarias}}

\author[G. Catanzaro and V. Ripepi]{G. Catanzaro$^{1}$\thanks{E-mail: gca@oact.inaf.it}, 
V. Ripepi$^{2}$
\\  
$^{1}$INAF-Osservatorio Astrofisico di Catania, Via S.Sofia 78, I-95123, Catania, Italy\\ 
$^{2}$INAF-Osservatorio Astronomico di Capodimonte, Via Moiariello 16, I-80131, Napoli, Italy\\
} 
 
\date{Accepted   Received ; in original form } 
 
\pagerange{\pageref{firstpage}--\pageref{lastpage}} \pubyear{2002} 
 
\def\LaTeX{L\kern-.36em\raise.3ex\hbox{a}\kern-.15em 
 T\kern-.1667em\lower.7ex\hbox{E}\kern-.125emX} 
 
\begin{document} 
 
\label{firstpage} 
 
\maketitle 
 
\begin{abstract} 

In this paper we present a detailed analysis of a sample of eight Am stars, four of them are in the {\it Kepler} field
of view. We derive fundamental parameters for all observed stars, effective temperature, gravity, rotational and radial
velocities, and chemical abundances by spectral synthesis method. Further, to place these stars in the HR diagram, 
we computed their luminosity.

Two objects among our sample, namely HD\,114839 and HD\,179458 do not present the typical characteristic of Am stars, 
while for the others six we confirm their nature. The behavior of lithium abundance as a function of the temperature with 
respect the normal A-type stars has been also investigated, we do not find any difference between metallic and normal
A stars. All the pulsating Am stars present in our sample (five out of eight) lies in the $\delta$~Sct instability strip, 
close to the red edge.

\end{abstract} 
 

 \begin{table*} 
\caption{Physical parameters estimated from photometry and
 parallaxes. The different columns show: (1) and (2) the HD number
 and an alternative name (if any) for the target star; (3) and (4)
 the $B$ and $V$ magnitudes adopted
 ($\sigma_B$,$\sigma_V$$\sim$0.020,0.015 mag, respectively); (5) the $K_\mathrm{s}$
 photometry from 2MASS ($\sigma_{K_\mathrm{s}} \sim$0.015 mag); (6)
 the $b-y$ colour ($\sigma_{b-y} \sim$ 0.01 mag); (7) the parallax
 \citep[after][]{leeuwen}; (8) the $E(B-V)$ values (the uncertainty is
 0.01 mag for the first four stars, 0.02 mag for the ramaining
 four objects; (9) the bolometric
 correction in the $V$ band \citep[after][]{bcp98}; (10) and (11) the
 $T_{\rm eff}$ estimated from $(V-K_\mathrm{s})$ and $uvby\beta$ photometry,
 respectively; (12) and (13) the $\log g$ estimated from $uvby\beta$
 photometry and Eq.~\ref{logg}, respectively.} 
\begin{center} 
\begin{tabular}{@{ }c@{ }lccccr@{ }c@{ }ccccc} 
\hline 
\hline 
\noalign{\medskip} 
HD  & Alt. Name &$B$& $V$ & $K_\mathrm{s}$& $b$$-$$y$ & $\pi$~~~~~~~~&$E$($B$$-$$V$) & $BC_V$ &$T^{(V-K_\mathrm{s})}_{\rm eff}$ & $T^{uvby\beta}_{\rm eff}$ & $\log g^{uvby\beta}$ &$\log g^{HIP}$\\ 
\noalign{\medskip} 
     &   &mag      & mag & mag  & mag &mas~~~~~~& mag                         &     mag     &        K                &           K               & cm/s$^{2}$    & cm/s$^{2}$        \\ 
(1)   &  (2)    & (3)         & (4)         &    (5)&(6) &    (7)~~~~~~~&  (8)&     (9)& (10)  & (11) & (12) & (13) \\ 
\noalign{\medskip} 
\hline 
\noalign{\smallskip}                                                 
104513  & DP UMa       &  5.475   &    5.207    &  4.553 &   0.171  &       29.18\,$\pm$\,0.75 &   0.00   &   0.092   &  7300$\pm$130  &   7380$\pm$120   &  4.09$\pm$0.14   &   4.18$\pm$0.07   \\             
113878  &              &  8.622   &    8.240    &  7.407 &   0.219  &       2.49\,$\pm$\,0.77  &   0.01   &   0.056   &  6930$\pm$130  &   7090$\pm$160   &  3.84$\pm$0.15   &   3.5$\pm$0.3     \\ 
114839  & NW Com       &  8.750   &    8.453    &  7.828 &   0.179  &       6.24\,$\pm$\,0.85  &   0.00   &   0.053   &  7420$\pm$130  &   7360$\pm$130   &  4.18$\pm$0.11   &   4.13$\pm$0.15   \\            
118660  & HR 5129      &  6.750   &    6.489    &  5.853 &   0.150  &       13.65\,$\pm$\,0.36 &   0.00   &   0.094   &  7340$\pm$130  &   7500$\pm$120   &  4.05$\pm$0.16   &   4.07$\pm$0.07   \\            
176843  & KIC\,9204718 &  9.053   &    8.745    &  8.043 &          &                          &   0.07   &   0.096   &  7610$\pm$210  &                  &                  &                   \\   
179458  & KIC\,9272082 &  9.154   &    8.954    &  8.568 &   0.074  &                          &   0.04   &   0.016   &  8150$\pm$230  &   8320$\pm$140   &  4.04$\pm$0.09   &                   \\  
187254  & KIC\,8703413 &  8.938   &    8.712    &  8.227 &          &                          &   0.04   &   0.095   &  7960$\pm$210  &                  &                  &                    \\
190165  & KIC\,9117875 &  7.821   &    7.495    &  6.747 &          &      10.06\,$\pm$\,0.43  &   0.06   &   0.083   &  7460$\pm$210  &                  &                  &        4.16$\pm$0.07           \\
\noalign{\smallskip} 
\hline 
\end{tabular} 
\end{center} 
\label{tabPhotometry} 
\end{table*}

\section{Introduction}
Among main sequence, A-type stars show a large variety of chemical
peculiarities. They are driven by several physical processes, such as
diffusion and/or magnetic field, just to quote some of them. All these
processes have the same factor in common, i. e. the very stable
radiative atmosphere which is the principal condition needed for
peculiarities to arise.

The metallic or Am stars are those whose Ca{\sc ii} K-line types
appear too early for their hydrogen line types, and metallic-lines
types appear too late, such that the spectral types inferred from the
Ca{\sc ii} K- and metal-lines differ by five or more spectral
subclasses. The marginal Am stars are those whose difference between
Ca{\sc ii} K- and metal-lines is less than five subclasses. The
commonly used classification for this class of objects include three
spectral types prefixed with {\it k}, {\it h}, and {\it m},
corresponding to the K-line, hydrogen-lines and metallic lines,
respectively. The typical abundances pattern show underabundances of
C, N, O, Ca, and Sc and overabundances of the Fe-peak elements, Y, Ba
and of the rare earths elements \citep{adelman97,fossati07}. The
presence of magnetic field has also been investigated but with null
result by \citet{fossati07}. The abundance of lithium in Am stars
compared to that observed in normal A-type stars, has been discussed
in the literature since the work of \citet{burkhart91}. They found that in
general Li abundance in Am stars is close to the cosmic value or even
lower in some case.

\citet{richer00} developed models of the structure and evolution of Am
stars in order to reproduce the observed chemical pattern of 28
elements. The most important improvement of these models has been the
introduction of turbulence as the hydrodynamical process competing
with atomic diffusion, in such a way that the resulting mixing reduces
the large abundance anomalies predicted by previous models, leading to
abundances which closely resemble those observed in Am stars.

Another open question in the framework of Am stars concerns the
pulsations in these objects. For many years it was thought that Am
stars did not pulsate, in accordance with the expectation that
diffusion depletes helium from the driving zone. Recently, intensive
ground-based \citep[SuperWASP survey]{smalley11} and space-based
\citep[{\it Kepler} mission]{balona11} observations have shown that
many Am/Fm stars do pulsate. \citet{smalley11}, for example, found
that about 169, 30 and 28 Am stars out of a total of 1600 show
$\delta$~Sct, $\gamma$~Dor or Hybrid pulsations \citep[see][for a
definition of these classes]{griga2010}.  These authors found also
that the positions in the Hertzsprung-Russel (HR) diagram of Am stars
pulsating as $\delta$~Sct are confined between the red and blue radial
fundamental edges, in agreement with \citet{balona11} and
\citet{catanzaro12}.

In this study we continue a programme devoted to determining
photospheric abundance pattern in Am stars by means of high resolution
spectra. Three Am stars have already been analyzed by us, namely:
HD\,178327 (KIC\,11445913) and HD\,183489 (KIC\,11402951) in
\citet{balona11}, and HD\,71297 in \citet{catanzaro13}, for which
fundamental astrophysical quantities, such as effective temperatures,
gravities and metallicities have been derived. The addition of these three
stars does not alter the homogeneity of our sample, since all of them have
been observed with the same instrumentation and the spectra
were reduced and analyzed with the same procedure that we will describe in
Sect.~\ref{obs}.
Such kind of studies are
crucial in order {\it i)} to put constraints on the processes
occurring at the base of the convection zone in non-magnetic stars and
{\it ii)} to try to define the locus on the HR diagram occupied by
pulsating Am stars.

With these goals in mind, we present a complete analysis of other
eight stars previously classified as Am stars.  Four of them belong to
the sample observed by the {\it Kepler} satellite \citep{balona11} and other four are Am
stars discovered to be pulsating from ground-based observations. For
our purposes high-resolution spectroscopy is the best tool principally
for two reasons, {\it i)} the blanketing due to the chemical
peculiarities in the atmospheres of Am stars alters photometric colors
and then fundamental stellar parameters based on them may not be
accurate \citep[see][and Sect.~\ref{comparison} for
details]{catanzaro12} and {\it ii)} the abnormal abundances coupled with
rotational velocity result in a severe line blending which
makes difficult the separation of the individual lines. Both problems
could be overcome only by matching synthetic and observed spectra.

For the confirmed Am stars we will compare the observed abundance with
the predictions of the models and we will place them on the HR diagram
by evaluating their luminosities.

\begin{table*}
\centering
\caption{Results obtained from the spectroscopic analysis of the
  sample of Am stars presented in this work. The different columns
  show: (1) identification; (2) effective temperatures; (3) gravity ($\log g$);
  (4) microturbolent velocity ($\xi$); (5) rotational velocity
  (v\,$\sin i$), (6) Heliocentric Julian Day of observation; (7)
  radial velocity (V$_{rad}$);  (8) indication of binarity (Y=binary;
  N=not binary; U=data insufficient to reach a conclusion); (9)
  indication of belonging to the Am star class (Y=Am; N=not Am); (10)
  indication of presence of pulsation 
  \citep[Y=pulsating; N=not pulsating; after][and references therein]{balona11}.
}
\begin{tabular}{cccccccccc}
\hline
\hline
\noalign{\medskip} 
 HD~~ &  T$_{\rm eff}$~~~~~~&  $\log g$ &  $\xi$ &v $\sin i$ ~~  & HJD  & V$_{rad}$~~~~  & Binary  & Am& Pulsating\\
           &            (K)      ~~~~~&                        & (km           s$^{-1}$)~ & (km s$^{-1}$)~~~&  (2450000.+)   &  (km           s$^{-1}$)    & & & \\
  (1)     &   (2)   &  (3)  & (4)  &(5)  &(6)  &(7)  &(8)  &(9)  &(10) \\
\noalign{\medskip} 
\hline    
\noalign{\smallskip} 
104513 & 7200\,$\pm$\,200 & 3.6\,$\pm$\,0.1 & 2.6\,$\pm$\,0.2 &72\,$\pm$\,7    & 5614.5119 & $-$3.12\,$\pm$\,1.71   &  Y    &  Y  & Y \\      
113878 & 6900\,$\pm$\,200 & 3.4\,$\pm$\,0.1 & 2.6\,$\pm$\,0.2 &65\,$\pm$\,6    & 5615.6761 & $-$2.78\,$\pm$\,1.53   &  Y    &  Y  & Y \\      
114839 & 7200\,$\pm$\,200 & 3.8\,$\pm$\,0.1 & 2.5\,$\pm$\,0.2 &65\,$\pm$\,7    & 5615.6388 & $-$2.87\,$\pm$\,1.72   &  U    &  N  & Y \\      
118660 & 7200\,$\pm$\,200 & 3.9\,$\pm$\,0.1 & 2.4\,$\pm$\,0.2 &100\,$\pm$\,10  & 5615.6490 & $-$1.70\,$\pm$\,0.25   &  N    &  Y  & Y \\    
176843 & 7600\,$\pm$\,150 & 3.8\,$\pm$\,0.1 & 2.7\,$\pm$\,0.2 &27\,$\pm$\,3    & 5696.6652 & $-$26.42\,$\pm$\,0.56  &  --   &  Y  & Y \\     
179458 & 8400\,$\pm$\,200 & 4.1\,$\pm$\,0.1 & 2.7\,$\pm$\,0.3 &75\,$\pm$\,7    & 5697.6103 & $-$15.37\,$\pm$\,0.57  &  --   &  N  & N \\        
187254 & 8000\,$\pm$\,150 & 4.1\,$\pm$\,0.1 & 2.5\,$\pm$\,0.2 &15\,$\pm$\,2    & 5697.6471 & $-$39.68\,$\pm$\,1.08  &  Y    &  Y  & N \\         
190165 & 7400\,$\pm$\,150 & 4.1\,$\pm$\,0.1 & 2.3\,$\pm$\,0.2 &58\,$\pm$\,6    & 5696.7086 & $-$7.45\,$\pm$\,0.45   &  U    &  Y  & N \\     
\noalign{\smallskip} 
\hline
\end{tabular}
\label{param}
\end{table*}

\section{Observation and data reduction}
\label{obs}

Spectroscopic observations of our sample of Am stars (see
Tab.~\ref{tabPhotometry} for the list of targets) were carried out
with the SARG spectrograph, which is installed at the {\it Telescopio
  Nazionale Galileo}, located in La Palma (Canarias Islands, Spain).
SARG is a high-resolution cross-dispersed echelle spectrograph
\citep{gratton01} that operates in both single-object and longslit
observing modes and covers a spectral wavelength range from 370~nm up
to about 1000~nm, with a resolution ranging from R = 29\,000 to
164\,000.

Our spectra were obtained in service mode in 2011, between February 21$^{th}$
and June 12$^{th}$, at R\,=\,57\,000 using two grisms (blue and yellow) and two 
filters (blue and yellow).  These were used in order to obtain a continuous 
spectrum from 3600 {\AA} to 7900 {\AA} with significant overlap in the wavelength 
range between 4620 {\AA} and 5140 {\AA}. We acquired the spectra for the stars with a 
signal-to-noise ratio S/N of at least 100 in the continuum. 

The reduction of all spectra, which included the subtraction of the bias frame, trimming, 
correcting for the flat-field and the scattered light, the extraction for the orders, and 
the wavelength calibration, was done using the NOAO/IRAF packages\footnote{IRAF is distributed 
by the National Optical Astronomy Observatory, which is operated by the Association of Universities 
for Research in Astronomy, Inc.}. The IRAF package {\it rvcorrect} was used to make the velocity 
corrections due to Earth's motion to transform the spectra to the barycentric rest frame. The 
radial velocities of our targets have been derived by cross correlating each observed spectrum 
with synthetic one. Results of this correlation, performed using the IRAF package {\it fxcor}, together 
with the heliocentric Julian day, have been reported in Tab.~\ref{param}.

\section{Physical parameters} 
\label{parameter} 
Temperatures and gravities for our sample stars have been derived by
spectral synthesis, as described in 
Sect.~\ref{parameters_from_spectroscopy}. In order to speed up the
iterative calculations, we needed starting values for both parameters,
that have been estimated from photometric calibrations, as described
in the following Sect.~\ref{param_phot}.  In the same section we
estimate the infrared excess and the bolometric corrections, needed to
compute stellar luminosities of our stars (see Sect.\ref{HR}).

\subsection{Parameters from photometry: $T_{\rm eff}$ and  $\log g$} 
\label{param_phot}
 
For five out of eight stars in our sample (HD\,104513, HD\,113878,
HD\,114839, HD\,118660, and HD\,179458) complete Str\"omgren-Crawford
$uvby\beta$ photometry is available \citep{hauck}. For the remaining 3
objects (HD\,176843, HD\, 187254, and HD\,190165), only Johnson
photometry is available, mainly in $BV$ filters. For these stars we
derived the Johnson $B,V$ magnitudes from Tycho ($B_T, V_T$)
photometry adopting the transformations into the standard system
provided by \citet{Bessell2000}. The same procedure was applied to all
the other stars for homogeneity. The resulting $B,V$ magnitudes are
listed in Tab.~\ref{tabPhotometry} (column 2 and 3).  In the
near-infrared, $JHK_\mathrm{s}$ photometry of good quality is present in the
2MASS catalogue \citep{2mass} for all the targets.
 
We adopted an updated version of the {\it TempLogG}\footnote{available
  through 
  http://www.univie.ac.at/asap/manuals/\\tipstricks/templogg.localaccess.html}
software \citep{rogers95} to estimate T$_{\rm eff}$ and $\log g$ by
using the calibrations present in the package, namely
\citet{balona84,moon85,md85,napi93,ribas97}.  In addition, we
considered the results by \citet{sk97} and \citet{heiter02} who
provided $uvby$ grids based on the Kurucz model atmospheres but with
different treatment of the convection. In particular, we used
\citet{sk97} grids built using \citet{cm91} convection treatment and
two choices for the grids\footnote{These grids are available on the
  NEMO site www.univie.ac.at/nemo/gci-bin/dive.cgi} by
\citet{heiter02}: i) standard mixing-length theory
(MLT)\footnote{defined as the ratio {\it $\alpha$=l/H$_p$} of
  convective scale length {\it l} and local pressure scale height {\it
    H$_p$}}; ii) the \citet{cgm96} treatment of the convection. For
each star, the different determinations T$_{\rm eff}$ and $\log g$
were comparable with each other and we decided to simply average
them. The result is shown in Table~\ref{tabPhotometry} (columns 9 and
10).

As for the reddening estimate, we have adopted different methods,
depending on the data available.  

\begin{itemize}

\item
For the five stars possessing $uvby\beta$
photometry, we used {\it TempLogG} to estimate the values of
$E(b-y)$, that were converted into $E(B-V)$ using the transformation
$E(B-V)=1.36\,E(b-y)$ \citep{cardelli89}.  

\item
We inspected the spectra af all our targets looking for the presence
of the interstellar lines Na{\sc i}  5890.0 \AA (D1) and K{\sc i} 7699
\AA. The equivalent widths (EWs) of these lines can be converted into $E(B-V)$
according to e.g. \citet{Munari1997}. As a result of this procedure,
the only measurable lines were Na{\sc i} in HD\,187254 (EW$\sim$140
m\AA) and K{\sc i} in HD\,179458 (EW$\sim$15 m\AA), corresponding to 
$E(B-V)$=0.04$\pm$0.02 mag for both stars. For the remaining objects the
interstellar lines were not measurable because they were too small 
(compatible with the almost zero absorption in the direction of 
HD\,104513,  HD\,113878,  HD\,114839, and HD\,118660 as derived from      
$uvby\beta$ photometry) or completely embedded into the photospheric
line. Is it worth noticing that for HD\,179458 the $uvby\beta$
photometry provided a different reddening estimate than that estimated
from K{\sc i}, and precisely $E(B-V)$=0.01$\pm$0.01 mag. Since we
judge that the \citet{Munari1997} calibration are reliable, for
HD\,179458 we decided to adopt the reddening evaluated from the
interstellar lines. 

\item
For the two remaining stars devoid of reddening estimate through the
aforementioned methods (namely, HD\,176843, HD\,190165), 
we adopted the tables by \citet{schmidt} in conjunction with
the spectroscopic $T_{\rm eff}$ and $\log g$ (see next section) to
estimate their instrinsic color $(B-V)_0$. A simple comparison with
the observed ones gives an estimate of the reddening for these
stars. 
\end{itemize}
The adopted reddening estimated are reported in Table~\ref{tabPhotometry}
(column 6).

To estimate a star's fundamental parameters from photometry and
parallax, we need to evaluate first the visual bolometric correction
BC$_V$. To this aim we adopted the models by \citet{bcp98} where it is assumed that 
M$_{\rm bol,\odot}$\,=\,4.74 mag. We interpolated their model grids
adopting the correct metal abundance that we derived in
Sect.~\ref{singleStarAbundances} as well as the values of $T_{\rm
  eff}$ and $\log g$ derived spectroscopically (see next section). The result of this procedure is
reported in Table~\ref{tabPhotometry} (column 7).

An additional photometric estimate of $T_{\rm eff}$ can be derived for
all the targets using the calibration $T_{\rm eff}$=$T_{\rm eff}$($(V-K_\mathrm{s})_0$,$\log g$ and [$Fe/H$])
published by e.g. \citet{masana06} or \citet{Casagrande2010}. Both
works give similar results and we decided to use \citet{masana06}'s
calibration for homogeneity with our previous papers \citep[e.g.][]{catanzaro11}.
As quoted above, the photometry in
$V$ and $K_\mathrm{s}$ is available from Tycho and 2MASS,
respectively. As for  $\log g$  and [$Fe/H$] we used the values from our 
spectroscopy. To de-redden the observed $(V-K_\mathrm{s})$ colours we adopted the reddening reported in 
Table\,\ref{tabPhotometry}, (column 4) using the relation $E(V-K_\mathrm{s}) =
2.8\,E(B-V)$ \citep[][]{cardelli89}. 
The resulting $T_{\rm eff}$ and the relative errors are reported in 
Table\,\ref{tabPhotometry} (column 8). 

Concerning $\log g$, it is possible to estimate with good accuracy this quantity independently 
from both spectroscopy and Str\"omgren photometry if the parallax is
known with sufficient precision (i.e. $\lesssim$ 10\%). As shown in
Tab.~\ref{tabPhotometry} (column 5), this is the case for three stars in our list,
namely  HD\,104513, HD\,114839, and  HD\,118660, whereas for 
HD\,113878 the error on the parallax is of the order of 30\%. 
To estimate $\log g$ we used the following expression:

\begin{eqnarray}
logg &= &4\log (T_{\rm eff} /T_{\rm eff,\odot}) + \log (M/M_\odot) +2 \log (\pi)
\nonumber \\
&&+0.4 (V+ BC_V + 0.26) + 4.44 
\label{logg}
\end{eqnarray}

\noindent
where the different terms of the above relationship have the usual meaning and
$M/M_\odot$ is the mass of the star in solar unit.
Before using Eq.~\ref{logg}, we have to evaluate the mass of the three
stars. This can be done by adopting the calibration 
mass--$M_V$ by \citet{malkov07} 
that was derived on the basis of a large sample of eclipsing binaries stars. 
Hence, by using our $M_V$ estimate discussed in Sect.~\ref{HR}, we evaluated $\log
(M/M_\odot)$=0.20, 0.24, and 0.15 dex with a common error of 0.05 dex
(dominated by the dispersion of the mass--$M_V$ relation) for
HD\,104513, HD\,114839, and  HD\,118660, respectively. For
HD\,113878 we obtained $\log (M/M_\odot)$=0.50$\pm$0.11 dex, being the 
error dominated by the large uncertainty on the parallax.  
Finally, the $\log g$ resulting from the above procedure are listed in column (11) of Table~\ref{tabPhotometry}.

\subsection{Atmospheric parameters from spectroscopy}
\label{parameters_from_spectroscopy} 
In this section we present the spectroscopic analysis of our sample of Am stars, in order to derive fundamental astrophysical
quantities, such as: effective temperatures, surface gravities, rotational velocities and chemical abundances. 

The approach used in this paper has been succesfully used in other papers devoted to this topics, see for instance 
\citet{catanzaro11,catanzaro12,catanzaro13}. In practice, the procedure used for our targets was to minimize the difference among 
observed and synthetic spectrum, using as goodness-of-fit parameter the $\chi^2$ defined as

\begin{equation}
\chi^2\,=\,\frac{1}{N}\sum \left(\frac{I_{obs} - I_{th}}{\delta I_{obs}}\right)
\end{equation}

\noindent
where N is the total number of points, I$_{obs}$ and I$_{th}$ are the intensities of the observed and computed profiles, respectively, 
and $\delta I_{obs}$ is the photon noise. Synthetic spectra were generated in three  steps. First, we computed LTE atmospheric models 
using the ATLAS9 code \citep{kur93,kur93b}. Second, the stellar spectra were then synthesized using SYNTHE \citep{kur81}. Third, the 
spectra were convolved for the instrumental and rotational broadenings.

We computed the $v \sin i$ of our targets by matching synthetic lines profiles from SYNTHE to a 
number of metallic lines. The Mg{\sc i} triplet at $\lambda \lambda$5167-5183 {\AA} was particularly useful for this purpose. The results
of these calculations are reported in Tab.~\ref{param}.

To determine stellar parameters as consistently as possible with the actual structure of the atmosphere, we performed the abundances 
analyses by the following iterative procedure: 

\begin{description}

\item{(i)} $T_{\rm eff}$ was estimated by computing the ATLAS9 model atmosphere which gave the best match 
between the observed H$_{\beta}$ and H$_{\delta}$ lines profile and those computed with SYNTHE. The models were 
computed using solar opacity distribution functions (ODF) and microturbulence velocities according to the 
calibration $\xi\,=\,\xi(T_{\rm eff},\log g)$ published by \citet{allende04}. For what concerns the treatment of 
convection, models cooler than 8000~K were computed using the classical MLT with fixed $\alpha$\,=\,1.25 \citep{castelli97}. 
The effects of different convection treatment on the Balmer lines profiles has already been investigated
in \citet{catanzaro13}, for the specific case study of HD\,71297. In that paper we concluded that theoretical 
profiles change according to the convection treatment, in the sense that the separation between the two profiles 
increases from the line core towards the wings. However, the maximum difference is very low, of the order of 
1.5\,$\%$, really indistinguishable at our level of S/N and for our resolving power. Since the star analyzed in 
that paper share the same classification (Am) of the targets presented here, and it has been observed with the 
same equipment (SARG@TNG) and in the same observing run, we are confident that the conclusions obtained in
\cite{catanzaro13} continue to apply also here.

These two Balmer lines are located far from the echelle orders edges so that it was possible 
to safely recover the whole profiles. The simultaneous fitting of two lines led to a final solution 
as the intersection of the two $\chi^2$ iso-surfaces. An important source of uncertainties arised from the 
difficulties in normalization as is always challenging for Balmer lines in echelle spectra. We quantified the 
error introduced by the normalization to be at least 100~K, that we summed 
in quadrature with the errors obtained by the fitting procedure. The final results for effective temperatures 
and their errors are reported in Tab.~\ref{param}.

\begin{figure} 
\centering
\includegraphics[width=9.5cm]{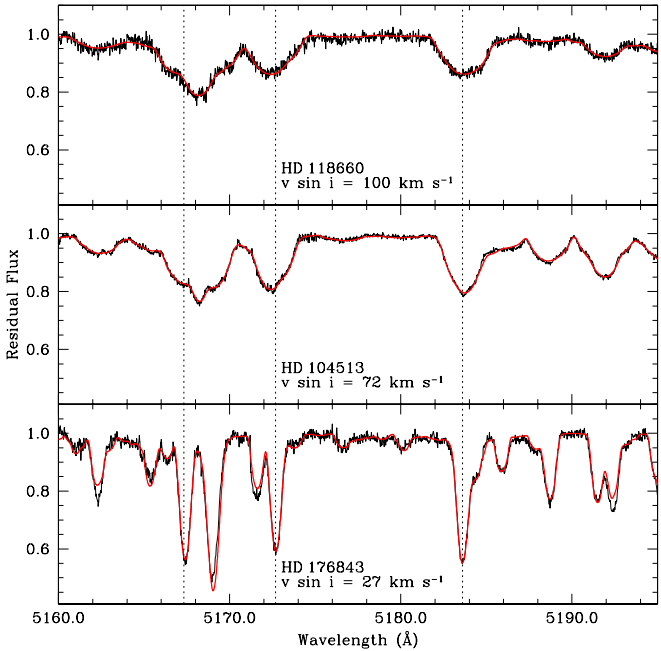} 
\caption{Observed Mg{\sc i} triplet superimposed with the corresponding synthetic spectra. The magnesium triplet lines are marked with
vertical dotted lines.} 
\label{mg} 
\end{figure} 

\begin{table*}
 \centering
  \caption{Abundances inferred for the Am stars of our sample. Values are expressed in the usual form $\log \frac{N_{\rm el}}{N_{\rm Tot}}$}
  \begin{tabular}{l@{ }c@{ }c@{ }c@{ }c@{ }c@{ }c@{ }c@{ }c@{ }c@{ }}
  \hline
  \hline
    &    HD\,104513         &     HD\,113878        &        HD\,114839	      &      HD\,118660	     &      HD\,176843       &      HD\,179458      &        HD\,187254      &       HD\,190165	    \\
\hline
Li  & $--$                  &  $-$8.90\,$\pm$\,0.10 &  $--$                 & $--$                 & $-$8.90\,$\pm$\,0.10 & $--$                 & $-$8.40\,$\pm$\,0.10 &  $-$9.00\,$\pm$\,0.10 \\                                                                                                                                           
C   & $-$3.46\,$\pm$\,0.11  &  $-$3.63\,$\pm$\,0.13 &  $-$3.71\,$\pm$\,0.17 & $-$3.50\,$\pm$\,0.13 & $-$3.70\,$\pm$\,0.11 & $-$3.14\,$\pm$\,0.10 & $-$3.48\,$\pm$\,0.11 &  $-$3.46\,$\pm$\,0.12  \\    
O   & $--$                  &  $--$                 &  $--$                 & $--$                 & $--$                 & $-$3.07\,$\pm$\,0.13 & $-$3.14\,$\pm$\,0.13 &  $--$                  \\    
Na  & $-$5.67\,$\pm$\,0.13  &  $-$5.57\,$\pm$\,0.14 &  $-$5.14\,$\pm$\,0.17 & $-$5.62\,$\pm$\,0.19 & $-$4.96\,$\pm$\,0.13 & $-$4.44\,$\pm$\,0.15 & $-$5.18\,$\pm$\,0.14 &  $-$5.41\,$\pm$\,0.13  \\    
Mg  & $-$4.50\,$\pm$\,0.15  &  $-$4.57\,$\pm$\,0.14 &  $-$4.10\,$\pm$\,0.06 & $-$4.59\,$\pm$\,0.10 & $-$4.40\,$\pm$\,0.15 & $-$4.32\,$\pm$\,0.17 & $-$3.86\,$\pm$\,0.14 &  $-$4.41\,$\pm$\,0.15  \\    
Al  & $--$                  &  $--$                 &  $--$                 & $--$                 & $-$5.20\,$\pm$\,0.14 & $-$4.04\,$\pm$\,0.14 & $-$4.81\,$\pm$\,0.09 &  $--$                  \\    
Si  & $-$4.49\,$\pm$\,0.11  &  $-$4.62\,$\pm$\,0.09 &  $-$4.46\,$\pm$\,0.10 & $-$4.41\,$\pm$\,0.12 & $-$4.41\,$\pm$\,0.12 & $-$4.21\,$\pm$\,0.10 & $-$4.11\,$\pm$\,0.12 &  $-$4.24\,$\pm$\,0.14  \\    
P   & $-$5.44\,$\pm$\,0.15  &  $--$                 &  $--$                 & $-$5.22\,$\pm$\,0.08 & $--$                 & $--$                 & $--$                 &  $--$                  \\    
S   & $-$4.51\,$\pm$\,0.12  &  $-$4.65\,$\pm$\,0.10 &  $-$4.52\,$\pm$\,0.18 & $-$4.36\,$\pm$\,0.12 & $-$4.61\,$\pm$\,0.11 & $-$4.00\,$\pm$\,0.12 & $-$4.44\,$\pm$\,0.11 &  $-$4.11\,$\pm$\,0.09  \\    
Ca  & $-$5.93\,$\pm$\,0.16  &  $-$5.72\,$\pm$\,0.13 &  $-$5.59\,$\pm$\,0.13 & $-$5.85\,$\pm$\,0.10 & $-$5.87\,$\pm$\,0.17 & $-$5.45\,$\pm$\,0.16 & $-$5.90\,$\pm$\,0.09 &  $-$6.28\,$\pm$\,0.15  \\    
Sc  & $-$9.70\,$\pm$\,0.15  &  $-$9.41\,$\pm$\,0.09 &  $-$9.53\,$\pm$\,0.17 & $-$9.00\,$\pm$\,0.08 & $-$9.17\,$\pm$\,0.10 & $-$8.50\,$\pm$\,0.16 & $-$8.93\,$\pm$\,0.07 &  $-$9.32\,$\pm$\,0.15 \\     
Ti  & $-$7.31\,$\pm$\,0.15  &  $-$7.26\,$\pm$\,0.12 &  $-$7.18\,$\pm$\,0.11 & $-$7.37\,$\pm$\,0.12 & $-$6.99\,$\pm$\,0.11 & $-$6.83\,$\pm$\,0.14 & $-$6.49\,$\pm$\,0.11 &  $-$6.90\,$\pm$\,0.15  \\    
Cr  & $-$6.19\,$\pm$\,0.19  &  $-$6.26\,$\pm$\,0.14 &  $-$6.29\,$\pm$\,0.18 & $-$6.15\,$\pm$\,0.03 & $-$5.87\,$\pm$\,0.12 & $-$6.25\,$\pm$\,0.17 & $-$5.56\,$\pm$\,0.10 &  $-$5.98\,$\pm$\,0.17  \\    
Mn  & $-$6.42\,$\pm$\,0.13  &  $-$5.89\,$\pm$\,0.09 &  $-$6.51\,$\pm$\,0.11 & $-$6.20\,$\pm$\,0.15 & $--$                 & $-$6.39\,$\pm$\,0.14 & $-$6.19\,$\pm$\,0.11 &  $-$6.12\,$\pm$\,0.11  \\    
Fe  & $-$4.20\,$\pm$\,0.18  &  $-$4.62\,$\pm$\,0.10 &  $-$4.60\,$\pm$\,0.17 & $-$4.10\,$\pm$\,0.12 & $-$4.07\,$\pm$\,0.15 & $-$4.30\,$\pm$\,0.13 & $-$3.84\,$\pm$\,0.07 &  $-$4.16\,$\pm$\,0.17  \\    
Co  & $--$                  &  $-$6.29\,$\pm$\,0.13 &  $-$6.45\,$\pm$\,0.14 & $-$6.37\,$\pm$\,0.06 & $--$                 & $-$5.83\,$\pm$\,0.08 & $-$5.87\,$\pm$\,0.07 &  $-$6.10\,$\pm$\,0.05  \\    
Ni  & $-$5.77\,$\pm$\,0.10  &  $-$5.61\,$\pm$\,0.12 &  $-$5.72\,$\pm$\,0.14 & $-$5.78\,$\pm$\,0.12 & $-$5.40\,$\pm$\,0.14 & $-$5.58\,$\pm$\,0.11 & $-$5.02\,$\pm$\,0.13 &  $-$5.45\,$\pm$\,0.15  \\    
Cu  & $--$                  &  $-$6.00\,$\pm$\,0.12 &  $--$                 & $--$                 & $-$7.22\,$\pm$\,0.10 & $--$                 & $-$6.98\,$\pm$\,0.10 &  $--$                  \\    
Zn  & $-$7.52\,$\pm$\,0.10  &  $--$                 &  $--$                 & $--$                 & $-$7.08\,$\pm$\,0.10 & $--$                 & $--$                 &  $--$                  \\    
Ge  & $--$                  &  $-$7.43\,$\pm$\,0.10 &  $--$                 & $--$                 & $-$7.57\,$\pm$\,0.06 & $--$                 & $--$                 &  $--$                  \\    
Sr  & $-$8.08\,$\pm$\,0.33  &  $-$8.33\,$\pm$\,0.10 &  $-$8.53\,$\pm$\,0.06 & $-$8.89\,$\pm$\,0.10 & $-$8.22\,$\pm$\,0.08 & $-$8.45\,$\pm$\,0.13 & $-$7.82\,$\pm$\,0.16 &  $-$7.76\,$\pm$\,0.13  \\    
Y   & $-$9.10\,$\pm$\,0.17  &  $-$9.10\,$\pm$\,0.09 &  $-$9.67\,$\pm$\,0.15 & $-$9.59\,$\pm$\,0.06 & $-$9.02\,$\pm$\,0.13 & $--$                 & $-$8.73\,$\pm$\,0.14 &  $-$9.06\,$\pm$\,0.12  \\    
Zr  & $--$                  &  $-$8.80\,$\pm$\,0.06 &  $--$                 & $-$8.77\,$\pm$\,0.08 & $-$8.76\,$\pm$\,0.19 & $--$                 & $-$8.69\,$\pm$\,0.07 &  $-$8.29\,$\pm$\,0.08  \\    
Ba  & $-$8.76\,$\pm$\,0.04  &  $-$8.56\,$\pm$\,0.14 &  $-$8.30\,$\pm$\,0.04 & $-$9.49\,$\pm$\,0.07 & $-$7.71\,$\pm$\,0.07 & $-$9.20\,$\pm$\,0.05 & $-$7.42\,$\pm$\,0.13 &  $-$8.46\,$\pm$\,0.14  \\    
 \hline
\end{tabular}
\label{abund}
\end{table*}

The surface gravity was estimated accordingly to the effective temperature of the star: for HD\,179458
and HD\,187254, i.e. the only stars of our sample hotter than 8000~K, we used the wings of Balmer lines as a 
diagnostic tool, while for the others, we derived $\log g$ from fitting the wings of broad lines of Mg{\sc i} 
triplet at $\lambda\lambda$~5167, 5172, and 5183 {\AA}, which are very sensitive to $\log g$ variations. As an example, we show in Fig.~\ref{mg} the fit for three stars of our sample, with
different rotational velocities. In practice, we have first derived the magnesium abundances through the narrow Mg{\sc i} lines at 
$\lambda \lambda$~4571, 4703, 5528, 5711~{\AA} (not sensitive to $\log g$), and then we fitted the wings of the triplet lines by fine tuning 
the $\log g$ value. To accomplish this task is mandatory to take into account very accurate measurements of the atomic parameters of the 
transitions, i.e. $\log gf$ and the radiative, Stark and Van der Waals damping constants. Regarding $\log gf$ we used the values 
of \citet{aldenius07}, Van der Waals damping constant is that calculated by \citet{barklem00} ($\log \gamma_{\rm Waals}\,=\,-7.37$), 
the Stark damping constant is from \citet{fossati2011} ($\log \gamma_{\rm Stark}\,=\,-5.44$), and the radiative damping constant is from 
NIST database ($\log \gamma_{\rm rad}\,=\,7.99$). 

The values of $\log g$, derived with this methods, have been checked by the ionization equilibrium between Fe{\sc i} lines (not
sensisitive to gravity change) and Fe{\sc ii} (very sensisitive to $\log g$). This procedure results in the final values reported in 
Tab.~\ref{param}.

Uncertainties in T$_{\rm eff}$, $\log g$, and $v \sin i$ were estimated by the change in parameter values
which leads to an  increases of $\chi^2$ by unity \citep{lampton76}.

\item{(ii)} As a second step we determine the stellar abundances by spectral synthesis. Therefore, we divide each of our spectra into 
several intervals, 50~{\AA} wide each, and derived the abundances in each interval by performing a $\chi^2$ minimization 
of the difference between the observed and synthetic spectrum. The minimization algorithm has been written in {\it IDL}
language, using the {\it amoeba} routine. We adopted lists of spectral lines and atomic parameters from \citet{castelli04}, who 
updated the parameters listed originally by \citet{kur95}.

\end{description}

For each element, we calculated the uncertainty in the abundance to be the standard 
deviation of the mean obtained from individual determinations in each interval of the 
analyzed spectrum. For elements whose lines occurred in one or two intervals only, 
the error in the abundance was evaluated by varying the effective temperature and 
gravity within their uncertainties given in Table~\ref{param}, 
$[ T_{\rm eff}\,\pm\, \delta T_{\rm eff}]$ and $[\log g\,\pm\,\delta \log g]$, and 
computing the abundance for $T_{\rm eff}$ and $\log g$ values in these ranges.
We found a variation of $\sim$0.1~dex due to temperature variation, while we did not
find any significant abundance change by varying $\log g$. The uncertainty in the temperature
is the main error source in our analyses.

\subsection{Comparison between astrophysical parameters derived by different methods}
\label{comparison}

It is useful to compare the values of $T_{\rm eff}$ and $\log g$ derived
spectroscopically (see Table~\ref{param}) with those obtained via
photometric methods (see Table~\ref{tabPhotometry}). 
Quantitatively, a weighted mean of the differences gives: 
$T_{\rm eff}^{\rm Spec}-T_{\rm eff}^{\rm (V-K_\mathrm{s})}$=$-$50$\pm$130 K;
$T_{\rm eff}^{\rm  Spec}-T_{\rm eff}^{\rm uvby\beta}$=$-$150$\pm$140 K.  
Similarly, for $\log g$:  
$\log g^{\rm Spec}$-$\log g^{uvby\beta}$=$-$0.25$\pm$0.25 dex;
$\log g^{\rm  Spec}$-$\log g^{HIP}$=$-$0.27$\pm$0.24 dex, or $-$0.15$\pm$0.11 dex if we exclude
the deviant star HD\,104513.

From an analysis of these results it appears that the 
$T_{\rm eff}^{\rm Spec}$ are in good agreement within the errors with
the $T_{\rm eff}$ estimated from ($V-K_\mathrm{s}$) colour,
whereas they are colder than $T_{\rm eff}^{\rm uvby\beta}$ by about
150 K, even if the significance of this value is only marginal ($\sim$
1 $\sigma$). 
Similarly, the $\log g^{\rm Spec}$ seems to be systematically smaller
than  $\log g^{uvby\beta}$ and, to a smaller extent, than $\log g^{HIP}$. In the first case the
discrepancy is not significant at 1$\sigma$ level. In the second case, 
with the exception of HD\,104513, there is agreement within
the errors. 

The star discrepant star HD\,104513 merits some further discussion. We
have carefully checked our photometric and spectroscopic data to
try to understand the $\sim$0.6 dex difference in logg between the
spectroscopic estimate and that derived from parallax. We have not
found errors in our procedures, and the star spectrum clearly
shows that HD\,104513 is evolved off the Main Sequence.  A possible
solution to the puzzle comes from the fact that HD\,104513 is a triple system
\citep[see e.g.][]{Tokovinin2008}. This occurrence could affect the
parallax measure and naturally explain the observed discrepancy.

The above results for $uvby\beta$ photometry are in agreement with
those by \citet{catanzaro12} who showed how the Str\"omgren indices
are correlated with effective temperature and $\log g$ and how they
are affected by blanketing in Am stars.  These authors concluded that effective
temperature can be reliably derived by Str\"omgren photometry, but
because the sensitivity of {\it (b-y)} to abundances, it is in general
higher of about 200\,K.  The situation is worst for the
gravities. Indeed, given the strong effect of blanketing on the {\it
  c$_1$} index, the gravities, and, in turn, the luminosities, are
completely unreliable.

\section{Chemical abundances}
\label{singleStarAbundances}

\begin{figure*}                                       
\includegraphics[width=8.8cm]{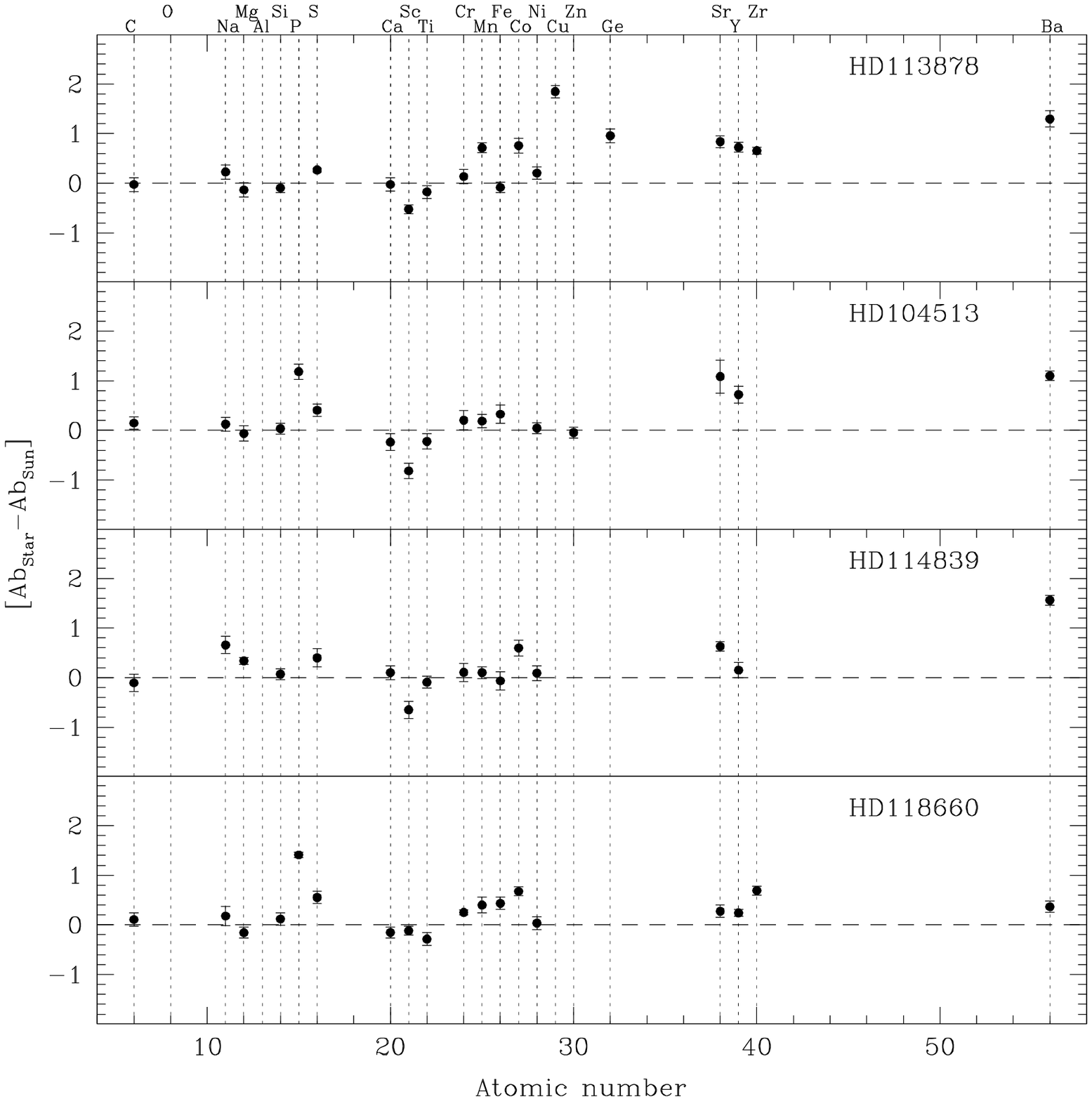}      
\includegraphics[width=8.8cm]{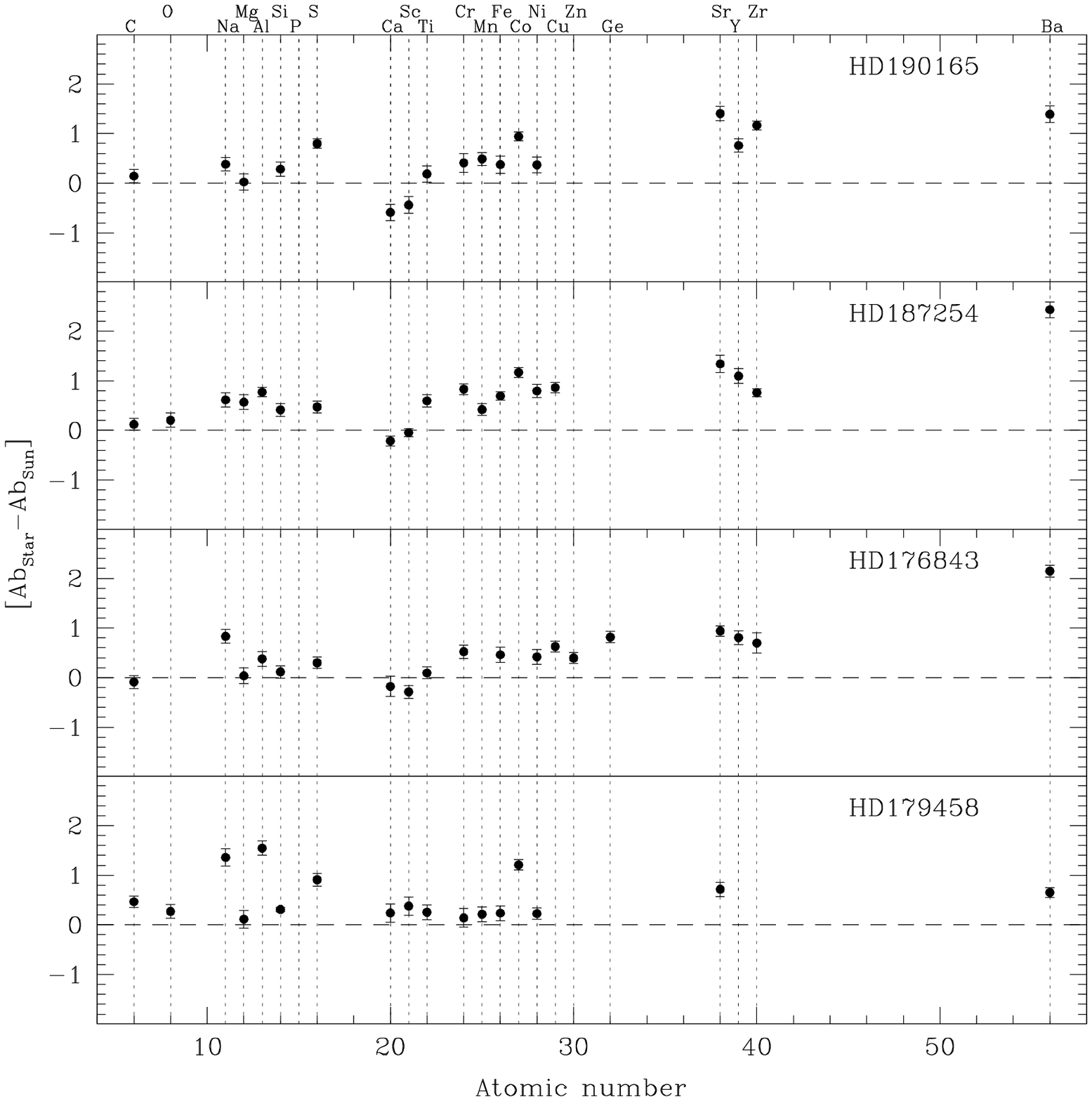} 
\caption{Chemical pattern for our targets, ordered by increasing effective temperature, from the coolest (top) to 
the hottest (bottom). Horizontal dashed line corresponds to solar abundance \citep{grevesse10}} 
\label{pattern1} 
\end{figure*}

In this section we present the results of the abundance analysis obtained for each star in our sample. 
The derived abundances and the estimated uncertainties, expressed as $\log \frac{N_{el}}{N_{Tot}}$, are reported in Tab.~\ref{abund}. 
The abundance patterns for each star, expressed in terms of solar values \citep{grevesse10}, are shown in Fig.~\ref{pattern1}. We also 
searched for binarity among our sample, combining 
our own measurements of radial velocity (reported in Tab.~\ref{param}) with those found in literature, when available. 

At the end of this section, we will discuss separately lithium abundance in Am stars with respect the normal A-type
stars.

\subsection{Individual stars}

\subsubsection{HD\,104513}
This star is known to be a metallic enhanced star since the pioneering work of \citet{morgan32}, who has noticed a strong Europium 
line at $\lambda$4129 {\AA}. \citet{cowley69},  by using metal spectral lines classified this star as A7 marginal metallic star, that 
is in agreement with that found later by \citet{hauck73}. This author, in his ``Catalogue of Am stars with known spectral types'', 
reported HD\,104513 to be an A7 from the Ca{\sc ii}~K line. \citet{abt75} found v$\sin i$\,=\,65\,$\pm$\,10~km~s$^{-1}$.

Radial velocity measurements have been found in the literature, \citet{abt85} published 23 velocities that are in agreement with the one
measured by us and reported in Tab.~\ref{param}. These velocities suggest a possible orbital motion, but since the amplitude is too low 
($\approx$\,5 km s$^{-1}$) compared to errors on each measurements, we cannot conclude anything on the binarity of this object.

HD\,104513 was the first marginal Am stars discovered to pulsate \citep{kurtz78}. He found indication of multiple periodicities in the 
$\delta$~Scuti regime, with periods ranging from 0.81 hr to 1.90 hr.

To our knowledge, this is the first extensive abundances analysis so
far published in the literature for HD\,104513. We estimated 
T$_{\rm eff}$\,=\,7100\,$\pm$\,200~K and $\log g$\,=\,3.6\,$\pm$\,0.1~dex,  that are typical for an F0/1 star, and
a v$\sin i$\,=\,72\,$\pm$\,7~km~s$^{-1}$ totally consistent with that published by \citet{abt75}. Moreover, we found moderate 
overabundances of about 1~dex for P, Sr, Y, and Ba, a slight overabundance of iron and iron peak elements and 
moderate underabundances of Ca and Sc, about 0.2~dex and 1~dex, respectively. Thus we confirm the classification of a marginal Am star, 
but from the Balmer and metallic lines we suggest it could be a star with a spectral type of F0/1.

\subsubsection{HD\,113878}
HD\,113878 was firstly classified as Am by \citet{olsen80}, who estimated its spectrum peculiarity on the basis of 
Str\"omgren photometric indices. Later on, this classification has been confirmed spectroscopically by \citet{abt84},
who define it as a {\it k}F1{\it h}F3V{\it m}F3 marginal Am star, because of its strong Sr{\sc ii} lines and weak 
Ca{\sc i} $\lambda$4026 {\AA} line. 

From the pulsational point of view this star has been intensitively studied in a series of paper by Joshi and collaborators.
\citet{joshi05}, in his photometric search for variability in Ap and Am stars, discovered this star to pulsate with a period 
of about 2.3 hours, which is typical of $\delta$ Scuti stars. This period has been refined later by \citet{joshi06}, who found
P\,=\,2.31 hr. Further observations carried out by \citet{joshi09} led the authors to conclude that HD\,113878
is an evolved star.

Regarding binarity, \citet{gouth06} reported $-$18.10\,$\pm$\,0.90~km~s$^{-1}$ and \citet{grenier97} measured $-$12.50\,$\pm$\,1.20~km~s$^{-1}$. 
Our value, $-$2.78\,$\pm$\,1.53~km~s$^{-1}$, confirmed the presence of a variation. 

Recently, \citet{casagrande11}, from spectroscopic observations, have derived the following astrophysical parameters for HD\,113878: 
T$_{\rm eff}$\,=\,7072\,$\pm$\,210~K, $\log g$\,=\,3.36, and [Fe/H]\,=\,0.74.

From our analysis, we found T$_{\rm eff}$\,=\,6900\,$\pm$\,200~K and $\log g$\,=\,3.4\,$\pm$\,0.1~dex,  
that are typical for an F1 evolved star, confirming both the results obtained by \citet{joshi09} and those 
from \citet{casagrande11}. Regarding the abundance pattern, we found a slight underabundance of scandium
of $\approx$\,0.5~dex, and a moderate overabundance of manganese, cobalt, germanium, strontium, yttrium, 
zirconium and barium, all ranging from 0.4 to $\approx$\,1~dex. A strong overabundance of copper, $\approx$\,1.8~dex,
has also been observed. This pattern confirms the classification of this star as a marginal Am star.

\subsubsection{HD\,114839}
Following \citet{hill76}, this object is reported in the ``General Catalogue of Ap and Am stars'' \citep{renson91} as an uncertain Am star.
\citet{pribulla09} carried out medium resolution (R\,=\,12000) spectroscopic observations at the David Dunlop Observatory, centered on 
the Mg{\sc i} triplet at $\lambda \lambda$5167--5184 {\AA}, from which they measured v$\sin i$\,=\,70~km~s$^{-1}$ and they concluded 
that it is a metallic line star of spectral type F4/5. \citet{balona11} reported a spectral type of {\it k}A5{\it h}F0{\it m}F3.

Only one measurement of radial velocity is reported in \citet{gouth06}: $-$5.60\,$\pm$\,1.40~km~s$^{-1}$. This value 
is in agreement with our own reported in Tab.~\ref{param}, at least within the experimental errors.

HD\,114839 has been discovered as hybrid pulsator by \citet{king06} by using space-based data carried out with the MOST satellite.
They identify 15 frequencies, of which 4 are in the range between 1 and 2.5 c/d, consistent with $\gamma$~Dor g-modes pulsations,
while the remaining are between 6.5 and 22 c/d, typical for $\delta$~Sct p-modes. 

For this star, we derived T$_{\rm eff}$\,=\,7100\,$\pm$\,200~K, $\log g$\,=\,3.8\,$\pm$\,0.1~dex,  and 
v$\sin i$\,=\,70\,$\pm$\,7~km~s$^{-1}$. These parameters led to a moderate ($\sim$0.5~dex) overabundances of Na, Mg, S, Co, and 
Sr and only a strong ($\sim$1.8~dex) overabundance of Ba. For what concerns the characteristic elements of the Am classification, we found
only a moderate underabundance of scandium, while other light and iron-peak elements are almost solar. Thus, in conclusion we cannot confirm
the Am peculiarity for this star.

A similar conclusion has been reached by \citet{hareter11}. They  performed an extensive spectroscopic study of HD\,114839 with the aim
to search for a link between the Am phenomenon and hybrid pulsators. Their effective temperature, surface gravity and rotational velocity are 
consistent with those derived in this study.

\subsubsection{HD\,118660}
\citet{barry70} was the first who noted marginal characteristic of Am phenomenology in the spectrum of HD\,118660. Later on,
\citet{cowley79} gave the first spectral classification relying on their H$\gamma$ spectrograms, denoting the star as a marginal A5m.

Two measurements of radial velocity have been reported in literature for HD\,118660, \citet{gouth06} ($-$1.7\,$\pm$\,2.9~km~s$^{-1}$) 
and \citet{wilson53} ($-$1.7~km~s$^{-1}$). Those values are in perfect agreement with our measured velocity, so we can confirm 
the absence of variability.

\citet{joshi06} discovered $\delta$~Scuti-like pulsations in this star, with a dominant period of about 1 hr and another prominent
period of about 2.52 hr.

To our knowledge, this is the first detailed abundance analysis performed for HD\,118660. Atmospheric parameters are: 
T$_{\rm eff}$\,=\,7200\,$\pm$\,200~K and $\log g$\,=\,3.9\,$\pm$\,0.1~dex,  and v$\sin i$\,=\,100\,$\pm$\,10~km~s$^{-1}$. 
Rotational velocity is consistent with the value reported by \citet{royer02} of 94~km~s$^{-1}$.
By using these values in our synthetic analysis, the most overabundance inferred was that of phosphorus of $\sim$1.5~dex. Moderate
overabundances in the range 0.2 - 0.6 dex have been found for S, Sc, iron-peak elements, Sr, Y, Zr and Ba. Solar to about $-$0.2 dex 
have been derived for other elements, including calcium and scandium. This result led us to conclude that HD\,118660 is a marginal Fm star.

This conclusion is corroborated by the work of \citet{charbo91}, who established a rotational velocity limit of 90~km~s$^{-1}$ above
which diffusion processes cannot cause Am peculiarities.

\subsubsection{HD\,176843}
HD\,176843 has been classified as {\it k}A3{\it m}F0, that is a marginal Am star, by \citet{floquet75}, but no studies are present in the
recent literature regarding its astrophysical parameters. 

Observed by the {\it Kepler} satellite, its periodogram has been presented firstly by \citet{balona11}, who discovered excess power
at two frequencies in the $\delta$~Sct domain, about at 34.4 c/d and 37.7 c/d. \citet{uytte11} classify this object as a binary
star with a $\delta$~Sct component. Unfortunately, we did not find any other measurements of radial velocity in literature, so 
we can not verify the possible binarity.

Even for this star, our study is the first ever reported in literature. Using the parameters we found, i.e. T$_{\rm eff}$\,=\,7600\,$\pm$\,150~K, 
$\log g$\,=\,3.8\,$\pm$\,0.1~dex and v$\sin i$\,=\,27\,$\pm$\,3~km~s$^{-1}$, we found slight underabundances of Ca and Sc, 
normal values for C, Mg, Si, and Ti, and overabundances for the heavier elements of about 0.5\,$\div$\,1~dex. Strong 
overabundance of Ba ($\sim$2~dex) have been observed, as well.

In conclusion this star shows the typical pattern of Am stars.

\subsubsection{HD\,179458}
The nature of this star has been debated in the past years, but in spite of this discussion, its classification is still doubtful.
\citet{MacRae52} noted its possible peculiar spectrum, but he did not give any details. Then the star was observed by 
\citet{floquet70}, which classified it as a normal A7 star. The uncertain nature is reported also in  the ``General Catalogue 
of Ap and Am stars'' \citep{renson91}. No measurements of radial velocity are present in literature.

Observed by {\it Kepler}, its periodogram does not show any sign of variability \citep{balona11}

Our study shows that HD\,179458 is an A4 main sequence star, with  T$_{\rm eff}$\,=\,8400\,$\pm$\,200~K, 
$\log g$\,=\,4.1\,$\pm$\,0.1~dex and v$\sin i$\,=\,75\,$\pm$\,7~km~s$^{-1}$. The most part of chemical elements 
observed in this star show overabundances, if compared with the respective solar values, from about 0.2~dex to 
about 1.5~dex. Besides its chemical pattern is far from the solar one, it is not
typical for Am stars, so we can conclude that HD\,179458 is not belonging to this class of peculiarity.

\subsubsection{HD\,187254}
Reported as a metallic star by \citet{mendoza74}, HD\,187254 has been then classified as {\it k}A2{\it m}F0 by \citet{floquet75}. 

Seven radial velocities have been reported by \citet{feren97}. Our measurement of radial velocity is compatible with those data,
so that we confirm the presence of an orbital motion since the amplitude is $\approx$~36 km s$^{-1}$, but we can not attempt for a search
of orbital parameters due to the lack of a sufficient number of data. 

From the pulsational point of view, this star has been studied by \citet{balona11} who analyzed the periodogram obtained with 
photometric data taken by the {\it Kepler} satellite. They concluded that it does not show any significant power excess in the $\delta$~Sct 
or $\gamma$~Dor range, though clear low-frequency variability is present. Some of this low-frequency variability may be of
instrumental origin as long-term trends in Kepler data are not fully corrected. However, intrinsic variability could arise as a result 
of rotational modulation, for example. While no Am star is known to vary in this way from ground-based observations, it cannot be ruled
out in {\it Kepler} photometry due to the extraordinary high precision.

Our study is the first ever detailed spectroscopic study, at least to our knowledge. From our spectrum we 
obtained: T$_{\rm eff}$\,=\,8000\,$\pm$\,150~K, $\log g$\,=\,4.1\,$\pm$\,0.1~dex and v$\sin i$\,=\,15\,$\pm$\,2~km~s$^{-1}$.
The only elements that appear to be solar are carbon and scandium, while a slight underabundance of $\approx$\,0.2~dex 
has been observed for calcium. Iron and iron-peak elements are slightly overabundant, as well light elements are.
Strong overabundances have been observed for Cu, Sr, Y, and Zr almost 1~dex, and for Ba, about 2.4~dex. Therefore 
no doubt that it is an Am star. 

\subsubsection{HD\,190165}
This star is known to belong to the Am group since the work of \citet{mendoza74}, who carried out multicolor photometry for a sample of
metallic stars. One year later, it was classified as {\it k}A2{\it m}F2 by \citet{floquet75}. Despite the fact that its nature has 
been known for a long time, both a detailed spectroscopic studies aimed at computing its chemical pattern and a measurement of
the rotational velocity for HD\,190165 are missing.

Regarding binarity, besides the two radial velocities reported in the literature are in agreement each other, 
v$_{\rm rad}$\,=\,$-$16.90~km~s$^{-1}$ \citep{gouth06,wilson53}, we found a discrepant value of v$_{\rm rad}$\,=\,$-$7.45\,$\pm$\,0.45~km~s$^{-1}$. 
In any case we can not make any conclusion about its variability.

{\it Kepler} observations have been analyzed by \citet{balona11} and, like the case of HD\,187254, they found only low-frequency variability.

From our spectrum we obtained T$_{\rm eff}$\,=\,7400\,$\pm$\,150~K and $\log g$\,=\,4.1\,$\pm$\,0.1~dex, and 
v$\sin i$\,=\,58\,$\pm$\,6~km~s$^{-1}$. The chemical pattern computed by using these parameters showed underabundances 
of about 0.5~dex for calcium and scandium, while heavy elements are all overabundant, from 0.4 dex for iron-peak 
elements to about 1.4~dex for barium.

In conclusion the Am nature of HD\,190165 is confirmed.

\subsection{Lithium abundance}

The lithium abundance in Am stars is a topic that has been discussed
in several papers in the recent literature. \citet{burkhart91} and
then \citet{burkhart05} concluded that, in general, lithium in Am
stars is close to the cosmic value of $\log N_{Li}/N_{Tot} \approx
-$9.04~dex, although a small fraction of them are Li
underabundant. \citet{fossati07} analysed a sample of eight Am stars,
belonging to the Praesepe cluster, in the range of temperature between
7000~K and 8500~K. By using the Li{\sc i} 6707 {\AA} line, they were able
to compute abundances that appears to be higher than the cosmic
value. \citet{catanzaro12} computed the abundance of lithium in the Am
star HD\,27411, deriving a value of $\log N_{Li}/N_{Tot} =
-$8.42\,$\pm$\,0.10, in agreement with the values reported by
\citet{fossati07}.

In this study we derived the lithium abundances for our Am stars (when
possible) and we compared them with those reported in various
literature sources for normal A-type stars.

\begin{figure}                                    
\includegraphics[width=9.5cm]{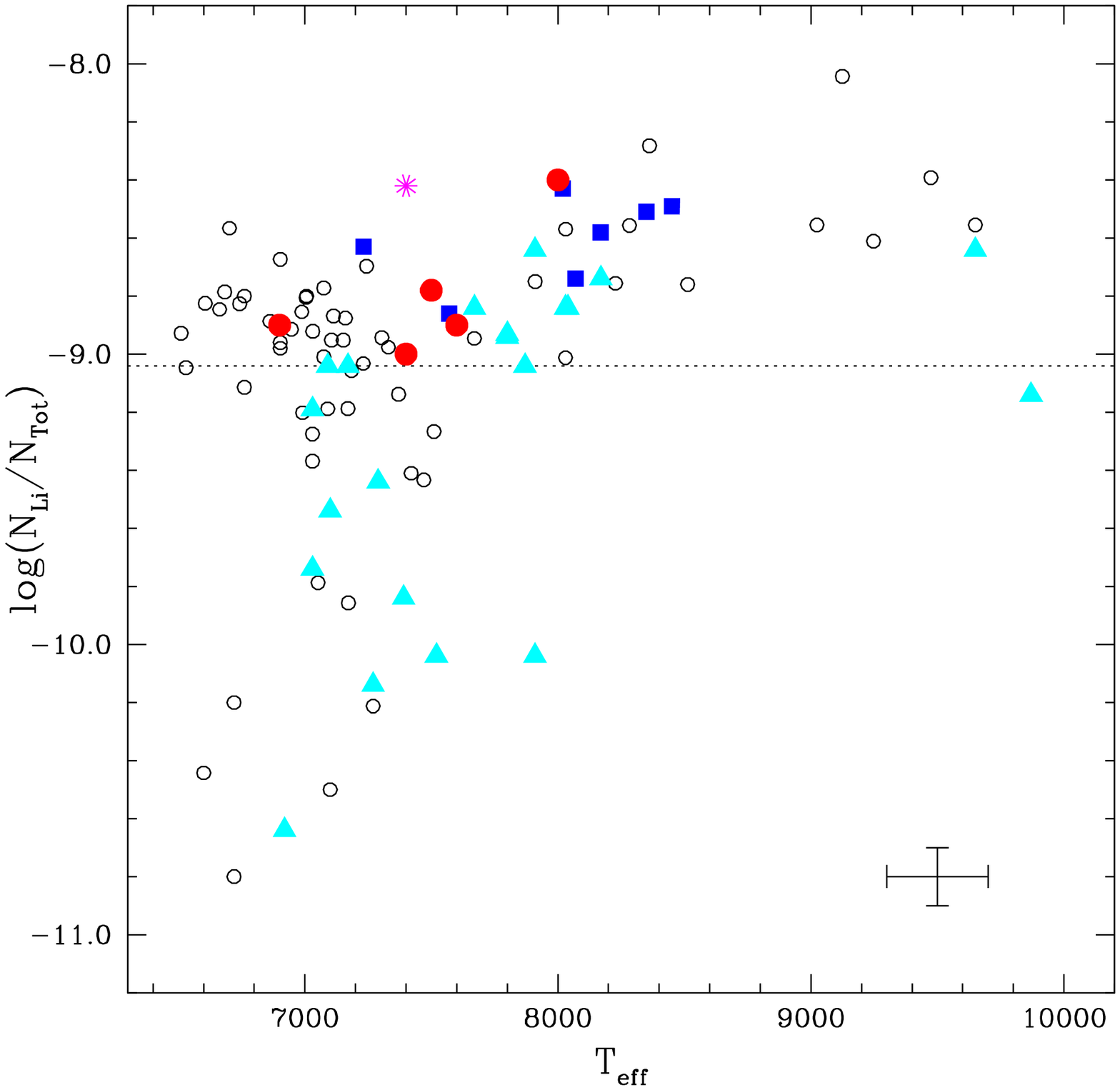} 
\caption{Lithium abundances plotted as a function of effective temperature. Filled symbols refer to Am
stars, in particular circles (red) represent our data, triangles (cyan) are from \citet{burkhart91}
and \citet{burkhart05}, squared (blue) are from \citet{fossati07}, and asterisk (magenta) are from
\citet{catanzaro12}. Opend circles refer to normal A-type stars taken from varius literature sources 
as outlined in the text. Typical errors are indicated in the bottom right corner of the plot.} 
\label{lithium} 
\end{figure}

To estimate the lithium abundance we applied the spectral synthesis
method to the Li{\sc i} 6707 {\AA} line, taking into account the
hyperfine structure \citep{andersen84}, as well. Due to the high
rotational velocity of some stars, we detected the line and than we
were able to compute the relative abundance for only five stars:
HD\,113878, HD\,176843, HD\,187254, HD\,190165 (see Tab.~\ref{abund}),
and HD\,71297 ($\log N_{Li}/N_{Tot} = -$8.78\,$\pm$\,0.11).  Lithium
abundances for these objects are shown (red filled circles) in
Fig.~\ref{lithium} as a function of the effective temperature.  For
comparison purposes we plotted in the same figures the lithium
abundances for various samples of Am stars. In particular we show with
cyan filled triangles the results by \citet{burkhart91} and
\citet{burkhart05} and with blue filled squares the data for Am
belonging to Praesepe cluster \citep{fossati07}.

With the aim of comparing the Lithium abundances in Am and normal A
stars, we computed the abundances for a sample of these latter objects
in two ways: i) converting the equivalent width of the Li{\sc i} 6707
{\AA} line taken from various sources: \citet{coupry92},
\citet{glaspey94}, \citet{balacha11} or ii) measured by us in spectra
available on the Elodie archive (Observatoire de Haute Provence).  For
homogeneity purpose, all the computations have been performed for all
the stars by using WIDTH9 \citep{kur81} applied to ATLAS9
models\footnote{For simplicity, since all the stars are on the main
  sequence, we set $\log g$\,=\,4 in the model computation}
\citep{kur93,kur93b}. These stars are listed in Tab.~\ref{normalA},
together with their effective temperatures, derived by using Str{\"o}mgren 
photometry as we described in Sect.~\ref{param_phot}, equivalent widths and Li abundances. 
The normal A-type stars are shown in Figure~\ref{lithium} with empty circles.

An inspection of Fig.~\ref{lithium} allows us to make some
reflections. First, the lithium abundance estimated in our sample of
Am stars is on average lower by $\approx$\,0.2 dex with respect to
that measured in the Am stars belonging to Praesepe cluster
\citep{fossati07}. Second, albeit our targets fall in the range of
effective temperatures of the so-called Li dip, a region of the
diagram T$_{\rm eff} - \log N_{Li}/N_{Tot}$ between the temperatures
6600~K to 7600~K, where lithium shows a
sudden drop of about 1.6 - 1.8 dex \citep{boesgard86}, none of them
present abundances lower than the cosmic value (see dotted line in
Fig.~\ref{lithium}). Third, it appears clear that there is no
difference between the peculiar and normal A-type stars. Even the Li
dip is present both in Am and in normal stars.

\begin{table}
\centering
\caption{Normal A-type stars and their Li abundances. For each star we reported its ID, effective temperature,
equivalent width of the Li{\sc i} 6707 {\AA} and the relative abundance expressed in the usual form $\log \frac{N_{\rm Li}}{N_{\rm Tot}}$.}
\begin{tabular}{rrrr}
\hline
\hline 
\noalign{\medskip} 
 HD    &  T$_{\rm eff}$  & EW  &  Abun \\
\noalign{\medskip} 
       &    K~   & m{\AA} & dex~   \\
\noalign{\medskip} 
\hline    
\noalign{\smallskip} 
739$^a$     & 6440 & 23.0 & -9.78 \\
2628$^a$    & 7270 &  3.2 &-10.21 \\
8829$^a$    & 7090 & 32.0 & -9.18 \\
30652$^a$   & 6480 & 20.0 & -9.90 \\
32537$^a$   & 7050 &  9.3 & -9.78 \\
33959$^a$   & 7670 & 25.0 & -8.94 \\
40136$^a$   & 7030 & 30.0 & -9.27 \\
43760$^a$   & 7170 & 29.0 & -9.18 \\
159834$^a$  & 8030 & 13.0 & -9.01 \\
205939$^a$  & 7910 & 27.0 & -8.75 \\
214994$^a$  & 9650 &  1.3 & -8.55 \\
24933$^b$   & 7510 & 16.0 &  -9.26  \\
79633$^b$   & 7100 &  2.0 & -10.50  \\
79765$^b$   & 7030 & 25.0 &  -9.36  \\
80390$^b$   & 6720 &  6.0 & -10.20  \\
106749$^b$  & 7330 & 42.0 &  -8.97  \\
112327$^b$  & 6720 &  2.0 & -10.80  \\
134360$^b$  & 7420 & 13.0 &  -9.41  \\
217602$^b$  & 7470 & 12.0 &  -9.43  \\
11636$^c$   & 8510 &  9.8 & -8.75 \\ 
25867$^c$   & 7030 & 59.0 & -8.92 \\ 
26322$^c$   & 7075 & 74.6 & -8.77 \\
27819$^c$   & 8280 & 22.4 & -8.55 \\
27962$^c$   & 9020 &  5.6 & -8.55 \\
37788$^c$   & 7230 & 38.3 & -9.03 \\
40932$^c$   & 8230 & 16.2 & -8.75 \\
50062$^c$   & 9120 & 13.0 & -8.04 \\
58946$^c$   & 7070 & 48.4 & -9.00 \\
89021$^c$   & 9250 &  2.9 & -8.61 \\
112412$^c$  & 6990 & 35.8 & -9.20 \\
132145$^c$  & 9470 &  3.3 & -8.39 \\
142908$^c$  & 6990 & 70.3 & -8.85 \\
150557$^c$  & 7150 & 48.0 & -8.95 \\
152598$^c$  & 7100 & 50.7 & -8.95 \\
166230$^c$  & 8360 & 33.5 & -8.28 \\
167858$^c$  & 7180 & 37.9 & -9.05 \\
177552$^c$  & 6950 & 65.5 & -8.91 \\
180777$^c$  & 7160 & 54.9 & -8.87 \\
217754$^c$  & 7170 &  7.2 & -9.85 \\
218396$^c$  & 7370 & 24.9 & -9.13 \\
219080$^c$  & 7300 & 40.9 & -8.94 \\
222603$^c$  & 8030 & 31.5 & -8.56 \\
\noalign{\smallskip} 
\hline
\end{tabular}
\\
$^a$ from \citet{coupry92}\\
$^b$ from \citet{glaspey94}\\
$^c$ from Elodie archive 
\label{normalA}
\end{table}     

\begin{table*}
\centering
\caption{Luminosities, distances and absolute visual magnitudes
  obtained from Eq.~\ref{eq} (columns 2--4) and from Hipparcos
  parallaxes (columns 5--7). See text for details.} 
\begin{tabular}{ccccccc}
\hline
\hline 
\noalign{\medskip} 
HD~~ & $\log L/{\rm L}_\odot$ & $D$ & $M^{0}_{V}$  & $\log L/{\rm L}_\odot$(HIP) & $D$(HIP) & $M^{0}_{V}$(HIP)\\
\noalign{\medskip} 
&     dex    &      pc                  & mag & dex &pc   &  mag     \\
(1)   &  (2)    & (3)            & (4)                       &    (5) &    (6)                &       (7)        \\
\noalign{\medskip} 
\hline    
\noalign{\smallskip} 
104513 &      1.25\,$\pm$\,0.14  &       54 $^{ +10  }_{-8}  $    & 1.51\,$\pm$\,0.16  &  0.85\,$\pm$\,0.03 &   34\,$\pm$\,1        &   2.53\,$\pm$\,0.07\\
113878 &      1.34\,$\pm$\,0.14  &      235 $^{ +42  }_{-36}$    & 1.33\,$\pm$\,0.16  &  1.81\,$\pm$\,0.27  & 402\,$\pm$\,124     &    0.17\,$\pm$\,0.67\\
114839 &      1.07\,$\pm$\,0.14  &      193 $^{ +35  }_{ -29}$   & 2.00\,$\pm$\,0.16  &   0.91\,$\pm$\,0.12  & 160\,$\pm$\,22     &     2.42\,$\pm$\,0.30\\
118660 &      0.98\,$\pm$\,0.14  &       72 $^{ +13  }_{  -11}$  & 2.20\,$\pm$\,0.16  &    1.00\,$\pm$\,0.03   & 73\,$\pm$\,2        &   2.16\,$\pm$\,0.07\\
176843 &      1.19\,$\pm$\,0.14  &      235 $^{+42  }_{  -35}$   & 1.66\,$\pm$\,0.15  & & & \\
179458 &      1.14\,$\pm$\,0.14  &      246 $^{ +43  }_{  -36 }$ & 1.87\,$\pm$\,0.15  & & & \\
187254 &      1.03\,$\pm$\,0.14  &      201 $^{ +35  }_{  -30 }$ & 2.06\,$\pm$\,0.15  & & & \\
190165 &      0.86\,$\pm$\,0.14  &      91 $^{ +16  }_{  -14 }$ & 2.51\,$\pm$\,0.15  &  0.94\,$\pm$\,0.05   & 99\,$\pm$\,4       &   2.32\,$\pm$\,0.13\\
\noalign{\smallskip} 
\hline
\end{tabular}
\label{paramLum}
\end{table*}  

\section{Position in the HR Diagram}
\label{HR}

In principle, the stellar parameters $\log g$ and $\log T_{\rm eff}$
determined in the previous section allow us to estimate
the luminosity of the investigated objects. In a previous paper
\citep{catanzaro11} we accomplished this task by interpolating 
the tables by \citet{schmidt}. However, it can be noticed that the space of
parameters $\log g$, $\log L/{\rm L}_\odot$, $\log T_{\rm eff}$ (spectral type)
adopted by \citet{schmidt}  is rather poorly sampled. To improve this situation we decided to use a
different calibration $\log L/{\rm L}_\odot$=$\log L/{\rm L}_\odot$($\log g$, $\log T_{\rm eff}$)
whose derivation is described in a different paper (Ripepi et al. in
preparation). Here we only show the final equation (valid in the intervals 3.2$<$$\log g$$<$4.7, 3.690$<$$\log T_{\rm eff}$$<$3.934):

\begin{eqnarray}
\log L/{\rm L}_\odot = (-15.46\,\pm\,0.34)+(5.185\,\pm\,0.080) \log T_{\rm eff}\\
 -(0.913\,\pm\,0.014) \log g~~{\rm rms=0.093~dex.} \nonumber
\label{eq}
\end{eqnarray}

Hence, we estimated the values of $\log L/{\rm L}_\odot$ for the eight
stars studied in this paper by inserting in the above equation the spectroscopically derived
values for $\log g$, $\log T_{\rm eff}$ as reported in
Tab.~\ref{param}. The result of this procedure is listed in
Tab.~\ref{paramLum} where we report in columns (2) to (4) the  $\log
L/{\rm L}_\odot$, the distance and the $M^{0}_{V}$, respectively. The
last two quantities were derived from the estimate of  $\log
L/{\rm L}_\odot$ by means of simple algebric passages and the
information included in Tab.~\ref{tabPhotometry}. 

As a check for these estimates, we derived the same quantities 
directly form the parallax measured by Hipparcos for four stars in our 
sample (see Tab.~\ref{tabPhotometry}). The results are
shown in column (5) to (7) of Tab.~\ref{paramLum}. A comparison
between column (2) and (5) reveals that the two independent
$\log L/{\rm L}_\odot$ estimates are in good agreeement within the
errors, with the exception of HD\,104513, which appears to be too
bright if luminosity is estimated by means of Eq.~\ref{eq}. We have
already discussed in Sect.~\ref{comparison} the possible origin of
this discrepancy. In any case, in the following 
we adopted the parallax-based $\log L/{\rm L}_\odot$ for HD\,104513, HD\,114839 
HD\,118660 and HD\,190165, whereas for HD\,113878 we preferred to
adopt the estimate from Eq.~\ref{eq}, given the large error on parallax.

\begin{figure}    
\centering                               
\includegraphics[width=9.5cm]{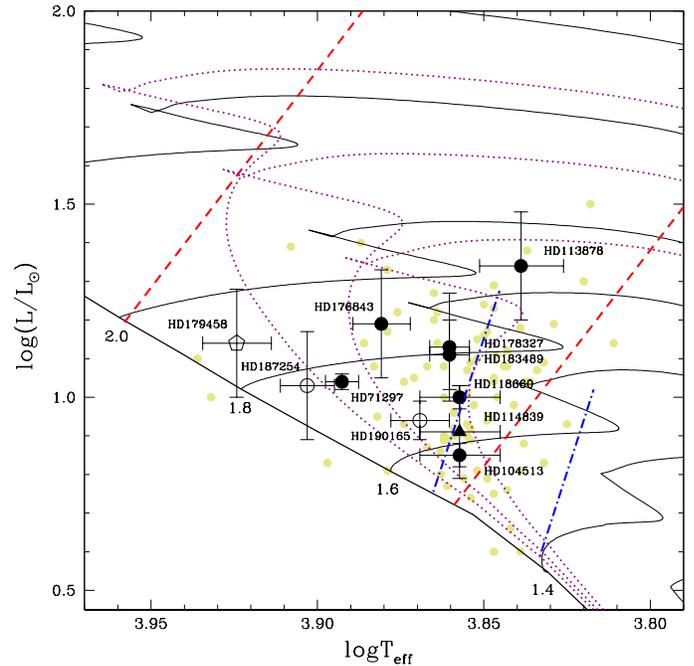} 
\caption{HR diagram for the eight stars investigated in this paper
 plus HD71297 \citep[after][]{catanzaro13}, HD\,178327 and HD\,183489 \citep[after][]{balona11}.
Note that the value of $\log L/{\rm L}_\odot$ of HD\,178327 was
artificially increased by 0.02 dex to avoid a complete overlap with HD\,183489.
Filled and empty circles show pulsating and non-pulsating Am stars,
respectively. The empty pentagon refers to a star that is neither Am nor
pulsating, whereas the filled triangles represent objects which are
pulsating but not Am. Small yellow filled circles show the pulsating
Am stars from the SuperWASP survey \citep{smalley11}
The red dashed lines shows the $\delta$\,Sct instability strip by \citet{breger98}; the blue 
dotted-dashed lines shows the theoretical edges of the $\gamma$\,Dor instability strip by 
\citet{guzik}. The evolutionary tracks (thin solid lines) for the
labelled masses as well as the ZAMS (thick solid line), and the isochrones for 0.5, 0.7 and 1.0 Gyrs (dotted lines)
are from the {\it BaSTI} database.} 
\label{figHR} 
\end{figure}

One of the aims of this paper is to try to constrain the locus occupied
by the pulsating Am star in the HR diagram. This is done in
Fig.~\ref{figHR} 
where we plotted the eight stars analysed in this paper ($\log T_{\rm
  eff}$ from column (2) of Table~\ref{param}; $\log L/{\rm L}_\odot$
from columns (2) or (5) of Table~\ref{paramLum}). In the same figure
we added the three pulsating Am stars analysed in our previous works, namely:
HD\,71297 \citep[after][]{catanzaro13}, HD\,178327 and HD\,183489
\citep[after][]{balona11}. Note that for the latter two stars, the
value of  $\log L/{\rm L}_\odot$ was recalculated. In particular, for 
HD\,183489, we used the Hipparcos parallax value
\citep[$\pi$=5.91$\pm$0.63;][]{leeuwen} to estimate its
luminosity, obtaining $\log L/{\rm L}_\odot$=1.11$\pm$0.09 dex. 
Unfortunately, Hipparcos did not observe HD\,178327. However, this
star appears to be a twin of HD\,183489,  showing exactly
the same $\log g$ or $\log T_{\rm eff}$ and chemical abundance
\citep[within the errors, see][]{catanzaro13} of this object. Hence, we decided to assign
to HD\,178327 the same luminosity of HD\,183489, but increasing the
error by 50\% to allow for the uncertainties in the stellar parameters
(i.e. $\log L/{\rm L}_\odot$=1.11$\pm$0.14 dex).
                          
To have an idea about the masses and ages of the investigated objects,
Fig.~\ref{figHR} shows the evolutionary tracks (solid lines) and the
isochrones (dotted lines) for 0.5, 0.7 and 1.0 Gyrs, respectively (the models, calculated
for Y=0.273, Z=0.0198, were taken from the {\it BaSTI}
database\footnote{http://albione.oa-teramo.inaf.it/}). 
We also show in the figure the comparison with the edges of the
$\delta$\,Sct \citep[after][]{breger98} and $\gamma$\,Dor
\citep[after][]{guzik} instability strips, respectively.

An analysis of the figure shows that only the cooler
part of the $\delta$\,Sct instability strip is occupied by the pulsating
Am stars investigated here, whereas no object falls in 
the region where only $\gamma$\,Dor pulsation is allowed. Only 
HD\,104513 (among the pulsating Am stars) lie in the region where 
both $\delta$\,Sct and $\gamma$\,Dor variability are excited. Moreover,
all the stars have an age between 0.5, 0.7 and 1.0 Gyrs.

For comparison purposes, Fig.~\ref{figHR} shows with small yellow filled
circles the location in the HR diagram of the pulsating Am stars found
by  the SuperWASP survey \citep{smalley11}. An inspection of the
figure reveals that our results are in perfect agreement with those
obtained by \citet{smalley11} on the basis of a larger sample: 
hot Am stars do not pulsate. This results is also valid for the object
observed with very high precision by the {\it Kepler}  satellite
\citep[see][]{balona11}. For the physical implication of this finding
we refer the reader to the quoted papers.

\section{Discussion and conclusion}

In this work we presented a spectroscopic analysis of a sample of 8
stars classified in literature as to belong to the class of the
metallic Am stars. The analysis is based on high resolution spectra
obtained at the {\it Telescopio Nazionale Galileo} with the SARG
spectrograph. For each spectra we obtained fundamental parameters such
as effective temperatures, gravities, rotational and radial
velocities, and we performed a detailed computation of the chemical
pattern, as well. To overcome the problem arising from blending of
spectral lines, we applyed the synthesis method by using SYNTHE
\citep{kur81} and ATLAS9 \citep{kur93} codes. The typical errors was
about 200~K for T$_{\rm eff}$, 0.1~dex for $\log g$, and a few
km~s$^{-1}$ for the v $\sin i$.

The values of T$_{\rm eff}$ and $\log g$ derived here have been used
to determine the luminosity of the stars and to place them on the HR
diagram.

According to our analysis, we ruled out two stars from the group of
the Am stars, namely: HD\,114839 and HD\,179458. The reasons are
different, HD\,114839 showed abundances almost solar in conten, while
HD\,179458 has a chemical pattern far from the solar one, but
nevertheless its peculiarity is not the one typical for Am stars.

All the observed stars lie in the $\delta$\,Sct instability strip next
to the red edge, in agreement with \citet{smalley11}
and \citet{catanzaro12}. 

In the scenario described by the diffusion models developed by
\citet{richer00}, stars in the range of temperature and age compatible
with those of our sample should have underabundances of about 0.1 to
0.3 dex for elements such as C, N, O, Na, Mg, K, and Ca, normal
abundance for Si and S, while Al, Ti, Cr, Mn, Fe, and Ni resulted
overabundant of about 0.1 to 0.8 dex.

For what that concern lithium, \citet{richer00} models predict
anomalies of $\approx$\,$-$0.2 dex with respect the cosmic value. For
our stars, in general we obtained abundances almost 0.2 dex over the
cosmic value, a result in agreement with the abundances found in the
Am star HD\,27411 \citep{catanzaro12} and in the Praesepe
cluster \citep{fossati07}. In conclusion we measured more lithium than
that predict by theory.

Recently, \citet{vick10}, in the context of the project to explore
various macroscopic processes which compete with atomic diffusion in
Am/Fm stars, computed a grid of models in which mass-loss has been
used instead of turbulence.  Those models predict at the side of Li
dip, where our objects lie, a smaller anomaly but still not sufficient
to explain our observations. As the authors suggested, it is likely
that more than one mechanism compete to diffusion, i. e. mass-loss in 
combination with turbulence, but at the moment is
not possible to conclude about one of this possibility.

In any case, our detailed abundance analysis can help theorist in
setting more constraints in their diffusion models.

\section*{Acknowledgments} 

We wish to thank our referee, Prof. Michael S. Bessell for his 
competent and useful review that helped us to improve the paper.\\
We would like to thank Dr. Luis Balona for his critical reading of our manuscript.\\
This research has made use of the SIMBAD database and VizieR catalogue access tool,
operated at CDS, Strasbourg, France. \\
This publication makes use of data products from the Two Micron All
Sky Survey, which is a joint project of the University of
Massachusetts and the Infrared 
Processing and Analysis Center/California Institute of Technology,
funded by the National 
Aeronautics and Space Administration and the National Science
Foundation.\\
This work has made use of BaSTI web tools.

\label{lastpage} 
 
\end{document}